\documentclass[pre%
,twocolumn%
,superscriptaddress%
,floats
,aps%
,amssymb%
]{revtex4}
\usepackage{amsmath}%

\begin{document}
\title{Stiffness Exponents for Lattice Spin Glasses in Dimensions $d=3,\ldots,6$} 
\date{\today}
\author{Stefan Boettcher}  
\email{sboettc@emory.edu}
\affiliation{Physics Department, Emory University, Atlanta, Georgia
30322, USA}  

\begin{abstract} 
The stiffness exponents in the glass phase for lattice spin glasses in
dimensions $d=3,\ldots,6$ are determined. To this end, we consider
bond-diluted lattices near the $T=0$ glass transition point
$p^*$. This transition for discrete bond distributions occurs just
above the bond percolation point $p_c$ in each dimension. Numerics
suggests that both points, $p_c$ and $p^*$, seem to share the same
$1/d$-expansion, at least for several leading orders, each starting
with $1/(2d)$. Hence, these lattice graphs have average connectivities
of $\alpha=2dp\gtrsim1$ near $p^*$ and exact graph-reduction methods
become very effective in eliminating recursively all spins of
connectivity $\leq3$, allowing the treatment of lattices of lengths up
to $L=30$ and with up to $10^5-10^6$ spins. Using finite-size scaling, data
for the defect energy width $\sigma(\Delta E)$ over a range of $p>p^*$
in each dimension can be combined to reach scaling regimes of about
one decade in the scaling variable $L(p-p^*)^{\nu^*}$. Accordingly,
unprecedented accuracy is obtained for the stiffness exponents
compared to undiluted lattices ($p=1$), where scaling is far more
limited. Surprisingly, scaling corrections typically are more benign
for diluted lattices. We find in $d=3,\ldots,6$ for the stiffness
exponents $y_3=0.24(1)$, $y_4=0.61(2), y_5=0.88(5)$, and
$y_6=1.1(1)$. The result for the upper critical dimension, $d_u=6$, suggest a mean-field value of $y_\infty=1$.
\hfil\break  PACS number(s): 
05.50.+q
, 64.60.Cn
, 75.10.Nr
, 02.60.Pn
.
\end{abstract} 
\maketitle
\section{Introduction}
\label{intro}
The stiffness exponent $y$ (often labeled $\theta$) is one of the most
fundamental quantities to characterize the low-temperature state of a
disordered spin system~\cite{F+H}. It provides an insight into the
effect of low-energy excitations of such a system~\cite{FH,BM1}. A
recent study suggested the importance of this exponent for the scaling
corrections of many observables in the low-temperature
regime~\cite{BKM}, and it is an essential ingredient to understand the
true nature of the energy landscape of finite-dimensional glasses
\cite{KM,PY2,PY}.

To illustrate the meaning of the stiffness exponent, one my consider
an ordinary Ising ferromagnet of size $L^d$ with bonds $J=+1$, which
is well-ordered at $T=0$ for $d>1$, having periodic boundary
conditions. If we make the boundary along one spatial direction
anti-periodic, the system would form an interface of violated bonds
between mis-aligned spins, which would raise the energy of the system
by $\Delta E\sim L^{d-1}$. This ``defect''-energy $\Delta E$ provides
a measure for the energetic cost of growing a domain of overturned
spins, which in a ferromagnet simply scales with the surface of the
domain. In a disordered system, say, a spin glass with an equal mix of
$J=\pm1$ couplings, the interface of such a growing domain can take
advantage of already-frustrated bonds to grow at a reduced or even
vanishing cost. Defect energies will be distributed with zero mean,
and the typical range, measured by the width of the distribution
$\sigma(\Delta E)$, may scale like
\begin{eqnarray}
\sigma(\Delta E)\sim L^y.
\label{yeq}
\end{eqnarray}
Clearly, it must be $y\leq d-1$, and a bound of $y\leq (d-1)/2$ has
been proposed for spin glass systems
generally~\cite{FH}. Particularly, ground states of systems with
$y\leq0$ would be unstable with respect to spontaneous fluctuations,
which could grow at no cost, like in the case of the one-dimensional
ferromagnet where $y=d-1=0$. Such a system does not manage to attain
an ordered state for any finite temperature. Conversely, a positive
sign for $y$ at $T=0$ indicates a finite-temperature transition into
an ordered regime while its value is a measure of the stability of the
ordered state. Furthermore, in a $d$-dimensional family of systems,
the marginal value $y_{d_c}=0$ provides the lower critical dimension
$d_c$ for such systems.

Accordingly, there have been many attempts to obtain the value of
stiffness exponents in finite-dimensional spin
glasses~\cite{PY,SY,Kirkpatrick,BC,BM,CB,Hy3d,Hd4,CBM,HBCMY,MKpaper},
using transfer matrix, optimization, or renormalization group
techniques. In the early days of spin-glass theory, it was soon argued
that $y<0$ for $d\leq2$ and $y>0$ for $d\geq3$~\cite{SY,BM}. Only
recently, though, the stiffness exponent for $d=2$, below the lower
critical dimension, has been improved to considerable accuracy,
$y_2=-0.282(2)$~\cite{HBCMY,CBM}. There has still been little progress
in the accurate determination of $y_3$ in the last 20 years, despite
significant increases in computational power. It's value is expected
to be small and positive, and so far has been assumed to be near
$y_3\approx0.19$~\cite{BM,Hy3d}, although there have been
investigations recently pointing to a larger value, such as
$0.23$~\cite{PY} or $0.27$~\cite{CBM}. In some sense, all of these
results are consistent, since they were obtained over exceedingly
small scaling windows, $L=6,\ldots,12$ at the best, and large errors
have to be assumed . In $d=4$ the only value reported to date has been
$y_4=0.64(5)$ using $L\leq7$~\cite{Hd4}.

In this paper we use numerical investigations of $\pm J$ spin glasses
on {\it dilute} lattices to obtain improved predictions for the
stiffness exponents in dimensions $d=3,\ldots,6$. First, we explore
such lattices near their bond-percolation transition $p_c$ to find a
separate transition $p^*>p_c$ into a $T=0$ spin-glass state, as
anticipated by Refs.~\cite{BF,MKpaper}. We find that $p^*$ becomes
ever closer to $p_c$ for increasing $d$, both scaling with
$1/(2d)$. Thus, near either transition, bond-diluted lattices have
spins with connectivities distributed near $2dp^*\approx1$. Such
sparse graphs can be effectively reduced with a set of {\it exact}
rules that eliminate a large fraction of spins, leaving behind a
small, compact remainder graph that is easier to optimize. The
increase in the scaling regime with lattice size, in combination with
finite-size scaling techniques, leads to much improved or entirely new
predictions for the stiffness exponents of low-dimensional
lattices. In particular, we find that $y_3=0.24(1)$, $y_4=0.61(2)$,
$y_5=0.88(5)$, and $y_6=1.1(1)$. Our value in $d=3$ is at the higher
end of most previous studies and amazingly close to (but distinct
from) the value obtained with the Migdal-Kadanoff approximation,
$y_3^{\rm MK}=0.25546(3)$~\cite{MKpaper}. The value for $d=4$ is
consistent with Ref.~\cite{Hd4} and quite below the Migdal-Kadanoff
value, $y_4^{\rm MK}=0.76382(5)$. The value for the upper critical
dimension, $d_u=6$, seems consistent with a recent mean-field prediction
of $y_{\infty}=1$, although that calculation was based on aspect-ratio scaling~\cite{AMY}.

In the following section, we discuss the observables that our numerical
experiments measure, in Sec.~\ref{algo} we describe the reduction
rules for low-connected spins and the optimization method use in this
study. Sec.~\ref{results} presents the results of the experiments for
the threshold $p^*$, the correlation-length exponent for the glass
transition, $\nu^*$, and the stiffness exponent $y$, in each
dimension. In Sec.~\ref{conclusion} we conclude with a discussion
regarding the $d$-dependence of $y$.

\section{Determining Stiffness Exponents}
\label{stiff}
To understand why the accurate determination of these stiffness
exponents is such a challenging task, it is important to appreciate
its complexity: Most numerical studies are based on sampling the
variance
\begin{eqnarray}
\sigma(\Delta E)=\sqrt{\left<\Delta E^2\right>-\left<\Delta
E\right>^2}
\label{sigmaeq}
\end{eqnarray}
of the distribution of defect energies $\Delta E$ obtained via
inverted boundary conditions (or variants thereof~\cite{CBM}), as
described above. Thus, for an Ising spin glass with periodic
boundaries, an instance of fixed, random bonds is generated, its
ground-state energy is determined, then all bonds within a hyperplane
have their sign reversed and the ground-state energy is determined
again. The defect energy is the often-minute difference between those
two ground state energies.  Then, many such instances of a given size
$L$ have to be generated to sample the distribution of $\Delta E$ and
its width $\sigma(\Delta E)$ accurately. Finally, $\sigma(\Delta E)$
has to be fitted to Eq.~(\ref{yeq}) for a sufficiently wide range of
$L$ in the asymptotic regime.

The most difficult part of this procedure, limiting the range of $L$
that can be achieved, is the accurate determination of the ground
state energies in the first place. While for $d\leq2$ efficient
algorithms exist to determine ground state energies exactly, and large
system sizes can be obtained~\cite{HBCMY,CBM}, for $d\geq3$ no such
algorithm exists: Finding ground states is known to be an NP-hard optimization problem~\cite{Barahona}
with the cost of any exact algorithm likely to rise faster than any
power of $L$.  There have been a variety of accurate measurements of
ground-state energies \cite{Pal,Hartmann_d3,eo_prl} using heuristic
methods. In these measurements small systematic errors in failing to
obtain a ground state tend to submerge beneath the statistical
error. In contrast, for the defect energy the extensive leading-order
contributions to the ground states are subtracted out, and such
systematic failings may surface to dominate any statistical
errors. Accordingly, system sizes that can be approximated with
heuristics may turn out to be far more limited than one may have
anticipated based on those previous studies.

To increase the range of system sizes $L$ without increasing the
optimization problem, we observe that a bond-diluted lattice will have
the same defect energy scaling as a fully connected lattice.  Above
the finite-size scaling window for bond-percolation near $p_c$, the
dominant cluster embedded on the lattice is a compact structure with
the same long-range properties of the fully connected
lattice. Similarly, the spin-glass problem defined on that cluster
should exhibit the same long-range behavior as the undiluted lattice
glass at $T=0$, their difference being of a short-range geometric
nature. Hence, for all bond densities $p$ above the scaling window of
the $T=0$ glass transition, Eq.~(\ref{yeq}) should be applicable. Yet,
a spin glass on a bond-diluted lattice in turn can be expected to be
less frustrated, up to the point that frustration fails to create
long-range correlated behavior. This is certainly the case below the
bond-percolation transition $p_c$, where any defects should remain
localized. Thus we focus on the regime somewhere above $p_c$, where the
system can exhibit spin-glass behavior but where we may take advantage
of the weakened frustration to optimize larger system sizes $L$.

As another feature of our new approach, the introduction of a new
control parameter, the bond density $p$, permits a finite size scaling
Ansatz. Combining the data for all $L$ {\it and} $p$ leads to a new variable which has the chance of exhibiting scaling over a
wider regime than $L$ alone. As has been argued in Ref.~\cite{BF},
we can make an Ansatz of
\begin{eqnarray}
\sigma(\Delta E)\sim {\cal Y}\, L^y
g\left[L\left(p-p^*\right)^{\nu^*}\right],
\end{eqnarray}
where ${\cal Y}\sim{\cal Y}_0(p-p^*)^f$ refers to the surface tension,
which must vanish for $p\to p^*$, and $g$ is a scaling function in the
new scaling variable, $x=L\left(p-p^*\right)^{\nu^*}$. The exponent
$\nu^*$ describes the divergence of the correlation length for the
transition into the ordered state at $p^*$. (In the Migdal-Kadanoff
approximation, it was found that $\nu^*$ is larger than $\nu$ of the
percolation transition~\cite{BF}.) Scale invariance at $p\to p^*$
dictates $f=y\nu^*$, and in terms of the scaling variable $x$ we have
\begin{eqnarray}
\sigma(\Delta E)\sim{\cal Y}_0\, x^y g(x).
\label{newyeq}
\end{eqnarray}

We will use the finite-size scaling relation in Eq.~(\ref{newyeq}) to
analyze our data in Sec.~\ref{defect}.  In the following section, we
describe the new algorithm for spin glasses on dilute lattices, which
at $T=0$ traces out many weakly connected spins to leave a much
reduced remainder graph which can be subsequently optimized by other
means.

\section{Reduction Algorithm for the Energies}
\label{algo}
We will describe the reduction algorithm for spin glasses on general
sparse graphs at $T=0$ in more detail elsewhere~\cite{eo_rg},
including its ability to compute the entropy density and overlap (see
also~\cite{MKpaper}). We focus here exclusively on the reduction rules
for the ground state energy.  We have used these reduction rules
previously for large three-connected Bethe lattices~\cite{eo_PRB}.
These rules apply to general Ising spin glass Hamiltonians
\begin{eqnarray}
H=-\sum_{<i,j>}\,J_{i,j}\,x_i\,x_j,\quad(x_i=\pm1),
\label{Heq}
\end{eqnarray}
with any bond distribution $P(J)$, discrete or continuous, on
arbitrary sparse graphs.  Here, we use exclusively a $\pm J$ bond
distribution, and bond-diluted hyper-cubic lattices in $d\geq3$. A
Gaussian or any other distribution with zero mean and unit variance is
expected to yield the same value of $y$~\cite{BKM}. Our preliminary
experiments with a Gaussian distribution have shown faster converging
averages at a given $L$, but more persistent scaling corrections for
large $L$.

The reductions effect both spins and bonds, eliminating recursively
all zero-, one-, two-, and three-connected spins and their bonds, but
also adding new bonds between spins which may or may not have been
connected previously. These operations eliminate and add terms to the
expression for the Hamiltonian in Eq.~(\ref{Heq}), leaving it {\it
form-invariant}. Offsets in the energy along the way are accounted for
by a variable $H_o$, which is {\it exact} for at $T=0$.

{\it Rule I:} An isolated spin, which does not contribute to the sum
in Eq.~(\ref{Heq}) at all, can be eliminated without changing that
sum.

{\it Rule II:} A one-connected spin $i$ can be eliminated, since its
state can always be chosen in accordance with its neighboring spin $j$
to satisfy the bond $J_{i,j}$, i.~e. in the only term in
Eq.~(\ref{Heq}) relating to $x_i$,
\begin{eqnarray}
x_i\,x_j\,J_{i,j}\leq\left|J_{i,j}\right|
\label{1coneq}
\end{eqnarray} 
we can always choose $x_i$ to saturate the bound, which is the
energetically most favorable state. With that, we adjust
$H_o:=H_o-|J_{i,j}|$ and eliminate the term $-J_{i,j}\,x_i\,x_j$ from
$H$.

{\it Rule III:} A double bond, $J_{i,j}^{(1)}$ and $J_{i,j}^{(2)}$,
between two vertices $i$ and $j$ can be combined to a single bond by
setting $J_{i,j}= J_{i,j}^{(1)}+J_{i,j}^{(2)}$ or be eliminated
entirely, if the resulting bond vanishes. This operation is very
useful, since it lowers the connectivity of $i$ and $j$ at least by
one. Particular to discrete bond distributions, there is a finite
probability that the two original bonds cancel each other
($J_{i,j}=0$), which may entirely {\it disconnect} $i$ and $j$ and
reducing their connectivity by two. (Double bonds are absent from the
original lattice but may arise via the recursive application of these
reduction rules.)

{\it Rule IV:} For a two-connected spin $i$, its two terms in
Eq.~(\ref{Heq}) can be rewritten
\begin{eqnarray}
J_{i,1}x_ix_1+J_{i,2}x_ix_2&=&x_i(J_{i,1}x_1+J_{i,2}x_2)\nonumber\\
&\leq&\left|J_{i,1}x_1+J_{i,2}x_2\right|\\
&=& J_{1,2}x_1x_2+\Delta H,\nonumber
\label{2coneq}
\end{eqnarray}
with
\begin{eqnarray}
J_{1,2}&=&\frac{1}{2}\left(\left|J_{i,1}+J_{i,2}\right|-\left|J_{i,1}-J_{i,2}\right|\right),\nonumber\\
\Delta H&=&\frac{1}{2}\left(\left|J_{i,1}+J_{i,2}\right|+\left|J_{i,1}-J_{i,2}\right|\right),\nonumber
\label{h0eq}
\end{eqnarray}
leaving the graph with a new bond $J_{1,2}$ between spin $1$ and $2$,
and acquiring an offset $H_o:=H_o-\Delta H$.

\begin{figure}[b!]
\vskip 1.1in \includegraphics{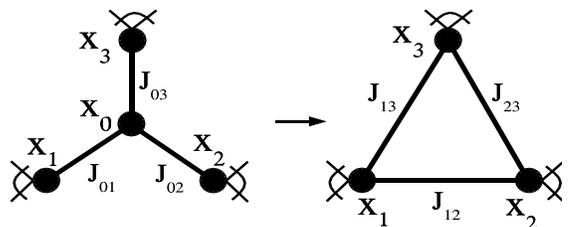}
\caption{Depiction of the ``star-triangle'' relation to reduce a
three-connected spin ($x_0$, center-left). The values for the new bonds on
the right are obtained in Eq.~(\protect\ref{3coneq}).  }
\label{startri}
\end{figure}

{\it Rule V:} A three-connected spin $i$ can be reduced via a
``star-triangle'' relation (see Fig.~\ref{startri}):
\begin{eqnarray}
&&J_{i,1}x_ix_1+J_{i,2}x_ix_2+J_{i,3}x_ix_3)\nonumber\\
&&~~~=x_i\left(J_{i,1}x_1+J_{i,2}x_2+J_{i,3}x_3\right)\nonumber\\
&&~~~\leq\left|J_{i,1}x_1+J_{i,2}x_2+J_{i,3}x_3\right|\\ 
&&~~~=J_{1,2}x_1x_2+J_{1,3}x_1x_3+J_{2,3}x_2x_3+\Delta H,\nonumber
\label{3coneq}
\end{eqnarray}
with
\begin{eqnarray}
&J_{1,2}=-A-B+C+D,\quad J_{1,3}=A-B+C-D,&\nonumber\\
&J_{2,3}=-A+B+C-D,\quad \Delta H=A+B+C+D,&\nonumber\\
&A=\frac{1}{4}\left|J_{i,1}-J_{i,2}+J_{i,3}\right|,\quad
B=\frac{1}{4}\left|J_{i,1}-J_{i,2}-J_{i,3}\right|,&\nonumber\\
&C=\frac{1}{4}\left|J_{i,1}+J_{i,2}+J_{i,3}\right|,\quad
D=\frac{1}{4}\left|J_{i,1}+J_{i,2}-J_{i,3}\right|.&\nonumber
\label{h1eq}
\end{eqnarray}
The bounds in Eqs.~(\ref{1coneq}-\ref{3coneq}) are saturated for the
right choice of the spin $x_i$ that links the terms together, thus
optimizing its alignment with the local field as is required when
the remaining graph takes on its ground state. In turn, for $T>0$ the
eliminated spin $x_i$ may not take on its own energetically most
favorable state to minimize the {\it free} energy of the configuration
instead. Hence, the reduction algorithm is exact only in determining the ground state.

Reducing four-~and higher-connected spins would lead to new bonds that
connect more than 2 spins, creating in general a hyper-graph with
multi-spin interaction terms. For instance, a term in $H$ connecting a
spin $\sigma_0$ to four other spins would be replace by one term
connecting all four, six terms mutually connecting the four neighbors
in all possible pairs, and an energy offset\footnote{As long as $H$
only contains terms connecting an even number of spins, the reduction
will preserve that evenness.}. While such a strategy may be useful, we
will confine ourselves here to reductions producing only new two-spin
interactions.

It is important that these rules are applied recursively and in the
given order. That is, one may only apply {\it Rule II} after there are
no more spins reducible by {\it Rule I}, apply {\it Rule III} only
after both, {\it Rule I} and {\it Rule II}, have been exhausted,
etc. And after the application of any higher rule, it needs to be
checked if structures have been generated to which any lower rule may
now apply. For example, the recursion may have generated a spin that
is two-connected, but via a double bond to a single other
spin. Applying {\it Rule IV} to that spin before {\it Rule III} would
lead to the other spin having a bond onto itself, a problematic
situation for which we have no rule. In any event, even if we had
provided more rules for all eventualities, it is still far more
efficient to first reduce the lowest connected spin at any one time.

After all these rules have been exhausted, the original lattice graph
is either completely reduced (which is almost certainly the case for
$p<p_c$), in which case $H_o$ provides the exact ground state energy
already, or we are left with a much reduced, compact graph in which no
spin has less than four connections. Note that bonds in the remainder
graph may have properties uncharacteristic of the original bond
distribution. For example, $\pm J$-bonds may have combined to bonds of
any integer multiple of $J$ (e.~g. via {\it Rule III}). Here, we
obtain the ground state energy of the reduced graph with the extremal
optimization heuristic~\cite{eo_prl}, which together with $H_o$
provides a very accurate approximation to the ground state energy of
the original diluted lattice instance. Clearly, we could have just as
well used other heuristics or exact methods to treat the remainder
graph.

\section{Numerical Results}
\label{results}
The following data was obtained during a window of about two months on
a cluster of 15 Pentium4 PC running at 2.4GHz with 256MB of RAM.

The runtime of the EO heuristic was fixed to grow as $(n/5)^3$ with the number $n$ of
spin variables in the remainder graph after the reduction had been applied. In Ref.~\cite{eo_prl} it was
found that typically $O(n^4)$ updates for instances up to $n\sim10^3$
were needed to obtain consistently reproducible ground state
energies. Since we are aiming at much larger statistics and typically smaller instances in the present
study, we opted for a more limited runtime. Instead, an adaptive
multiple restart system was used such that for each instance at least
3 runs from fresh random initial spin configurations were
undertaken. If a new best-so-far configuration is found in run $r$, at
least a total of $2r$ restarts would be applied to these more
demanding instances~\cite{BoPe3}. For instances with $n>700$
apparent inaccuracies in sampling the {\it difference} between ground
state energies, $\Delta E$, become noticeable.

For highly connected graphs with few spins to reduce, local search
with the EO heuristic dominated by far the computational time. Our
implementation of the reduction algorithm, originally conceived with
$d=3$ lattices with up to $L=30$ in mind, started contributing
significantly to the computational cost for instances with $L^d>10^5$,
hence most noticeably in the study of $p^*$ in $d=5$ and 6.

\subsection{Determination of $p^*$}
\label{pstar}
In Ref.~\cite{BF} it was shown that spin glasses on diluted lattices
may possess a distinct critical point $p^*$ in their bond fraction,
which arises from the (purely topological) percolation threshold $p_c$
of the lattice in conjunction with a discrete distribution of the bond
weights $P(J)$. Clearly, no long-range correlated state can arise
below $p_c$. A critical point distinct from percolation, $p^*>p_c$,
emerges when such an ordered state above $p_c$ remains suppressed due
to collaborative effects between bonds~\cite{BF} (see {\it Rule III}
in Sec.~\ref{algo}). Just above $p_c$, the infinite bond-cluster is
very filamentary and may easily be decomposed into finite components
through such collaborative effects, involving a small number of bonds
along narrow ``bridges'' between those components. Thus, to observe
the onset of glassy properties on a dilute lattice, we have to cross
another threshold $p^*\geq p_c$ first. In Ref.~\cite{MKpaper}, we were
able to locate $p^*$ for the Migdal-Kadanoff lattice in accordance
with theory~\cite{BF} by using the defect energy scaling from
Eq.~(\ref{yeq}): For all $p>p^*$ the stiffness exponent $y$ eventually
took on its $p=1$ value, while for any $p<p^*$ defect energies
diminished rapidly for increasing $L$.

\begin{table}
\caption{List of the bond-density thresholds on hyper-cubic lattices
for percolation $p_c$ (taken from Ref.~\protect\cite{Hughes}) and for
the $T=0$-transition into a spin glass state, $p^*$, as determined
from Figs.~\protect\ref{pstarfig}. The values of $p^*$ are dependent
on the bond distribution which is $\pm J$ here.}
\begin{tabular}{r|ll}
\hline\hline 
$d$      & $p_c$           & $p^*$           \\ 
\hline 
3        & 0.2488          & 0.272(1)         \\ 
4        & 0.160130        & 0.1655(5)        \\ 
5        & 0.118174        & 0.1204(2)        \\ 
6        & 0.0942          & 0.0952(2)        \\ 
\hline\hline
\end{tabular}
\label{pstartable}
\end{table}

\begin{figure}
\vskip 8in 
\includegraphics{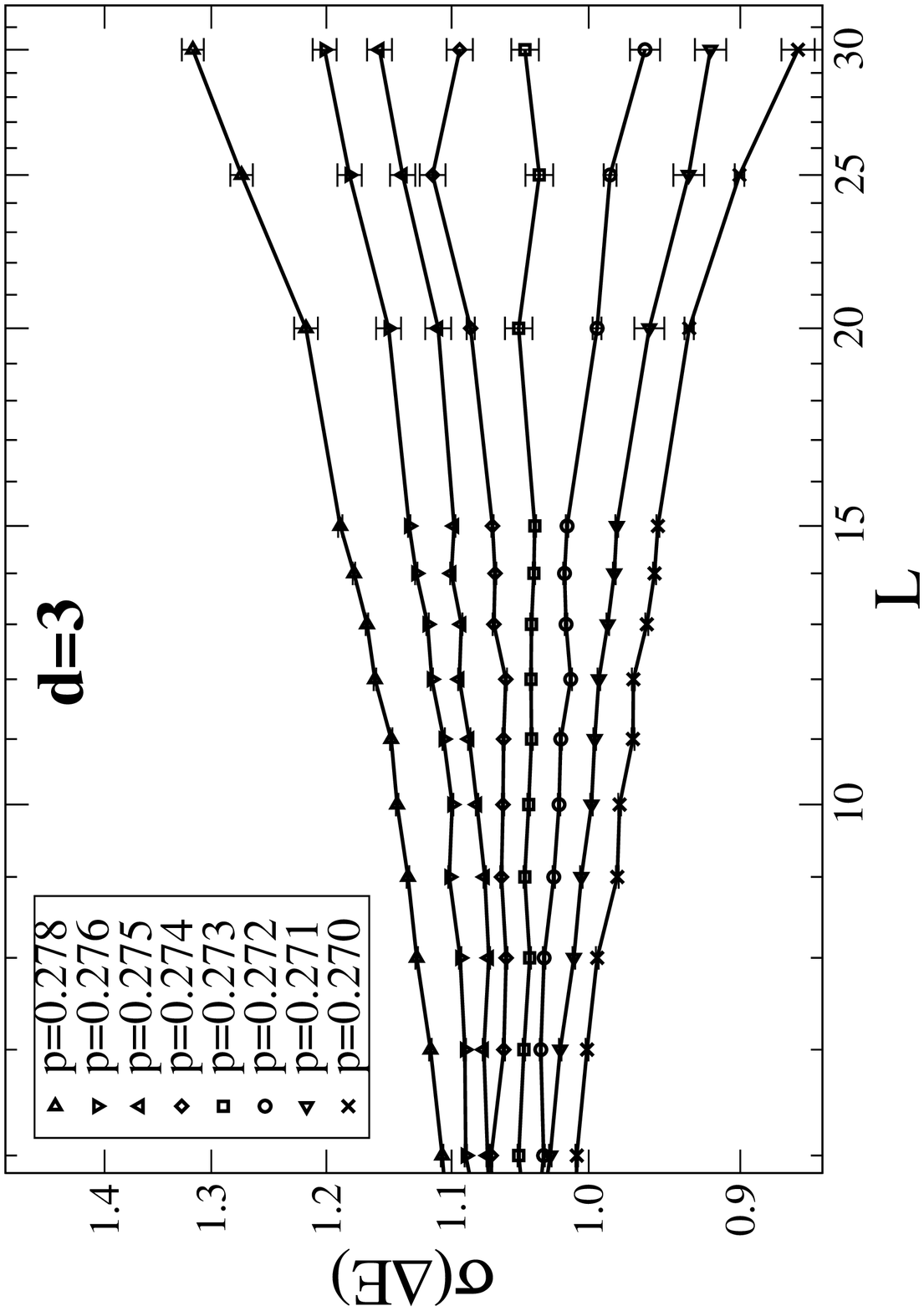}
\includegraphics{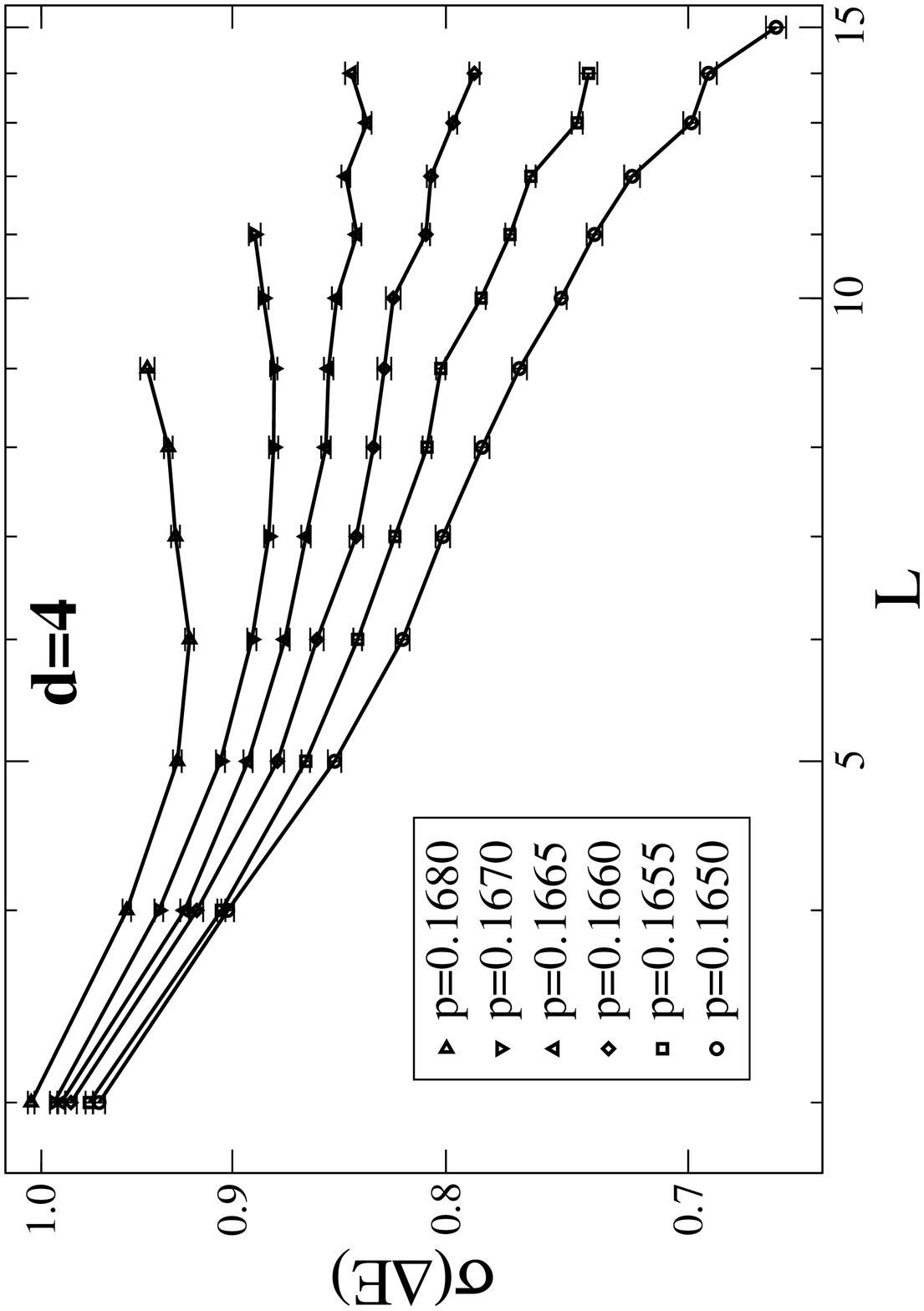}
\includegraphics{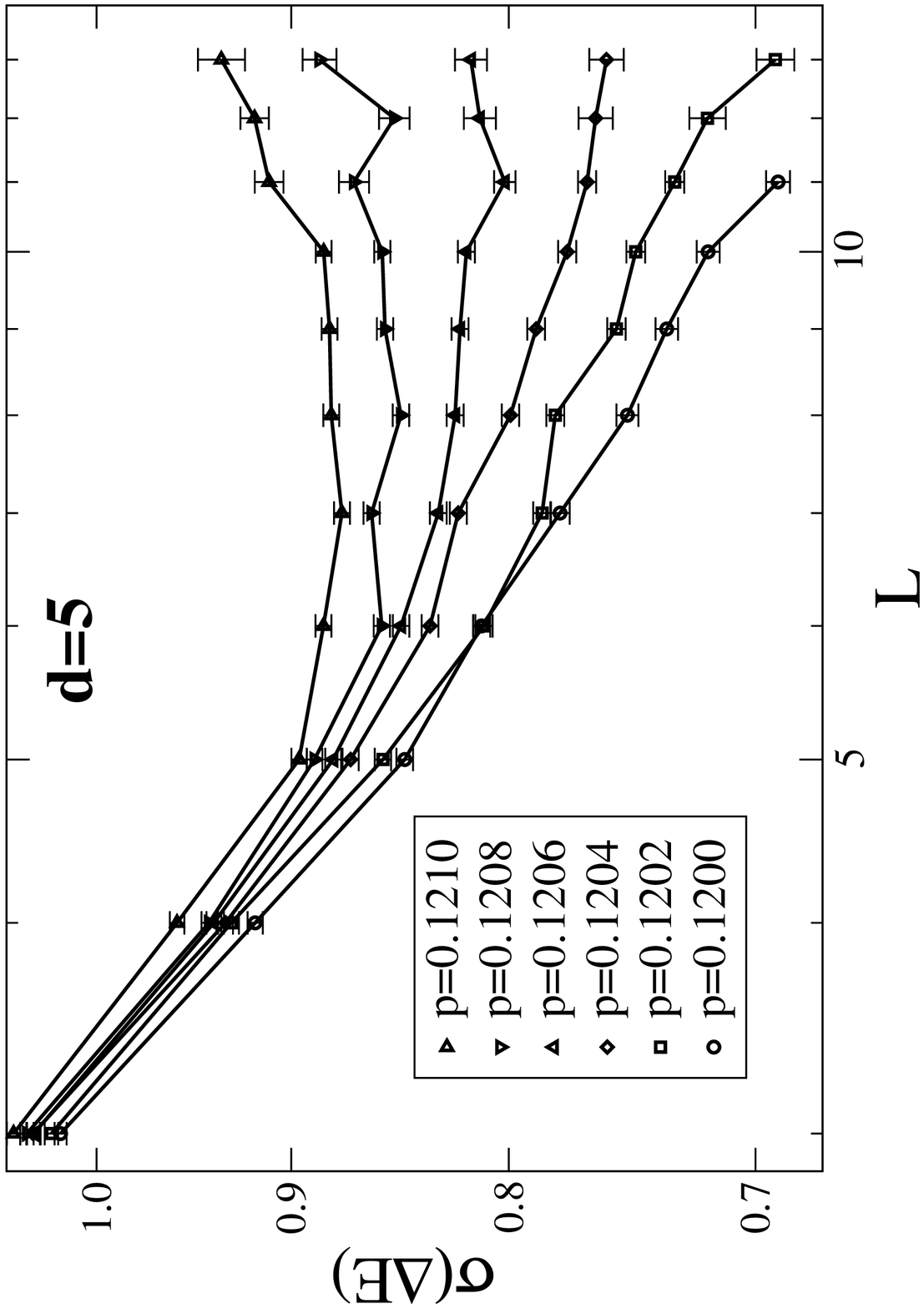}
\includegraphics{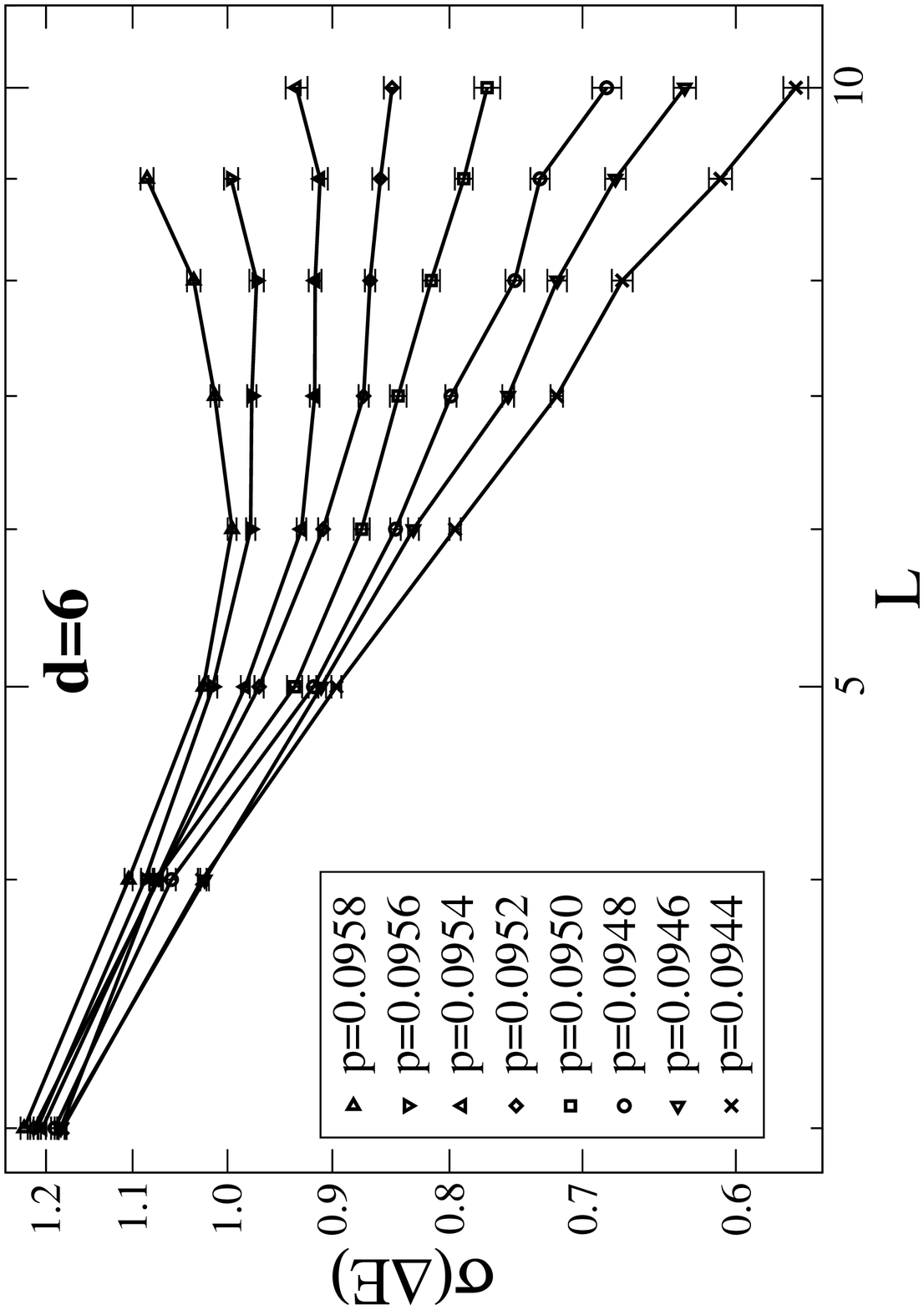}
\caption{Plot on a logarithmic scale of the variance $\sigma(\Delta
E)$ of the defect energy as a function of systems size $L$ for various
bond fractions $p>p_c$ in $d=3$ to 6. In each case, $\sigma(\Delta E)$
drops to zero rapidly for increasing $L$ at smaller $p$, but turns
around and rises for larger $p$, indicative of a nontrivial glassy
state at low $T$. Near $p^*$, $\sigma(\Delta E)$ undergoes ever longer
transients. The values for the thresholds $p^*$ as suggested by the
plots are listed in Tab.~\protect\ref{pstartable}.}
\label{pstarfig}
\end{figure}

In each dimension, we have run the above algorithm on a large number
of graphs (about $10^5-10^6$ for each $L$ and $p$) for $p$ increasing
from $p_c$ in small steps. For each given $p$, $L$ increased until it
seemed clear that $\sigma(\Delta E)$ would either drop or rise for
good. In this way, we bracket-in $p^*$, as shown in
Figs.~\ref{pstarfig}. Both, the bond-percolation thresholds $p_c$,
taken from Ref.~\cite{Hughes}, and our results for $p^*$ are listed in
Tab.~\ref{pstartable}.

\begin{figure}
\vskip 2.1in 
\includegraphics{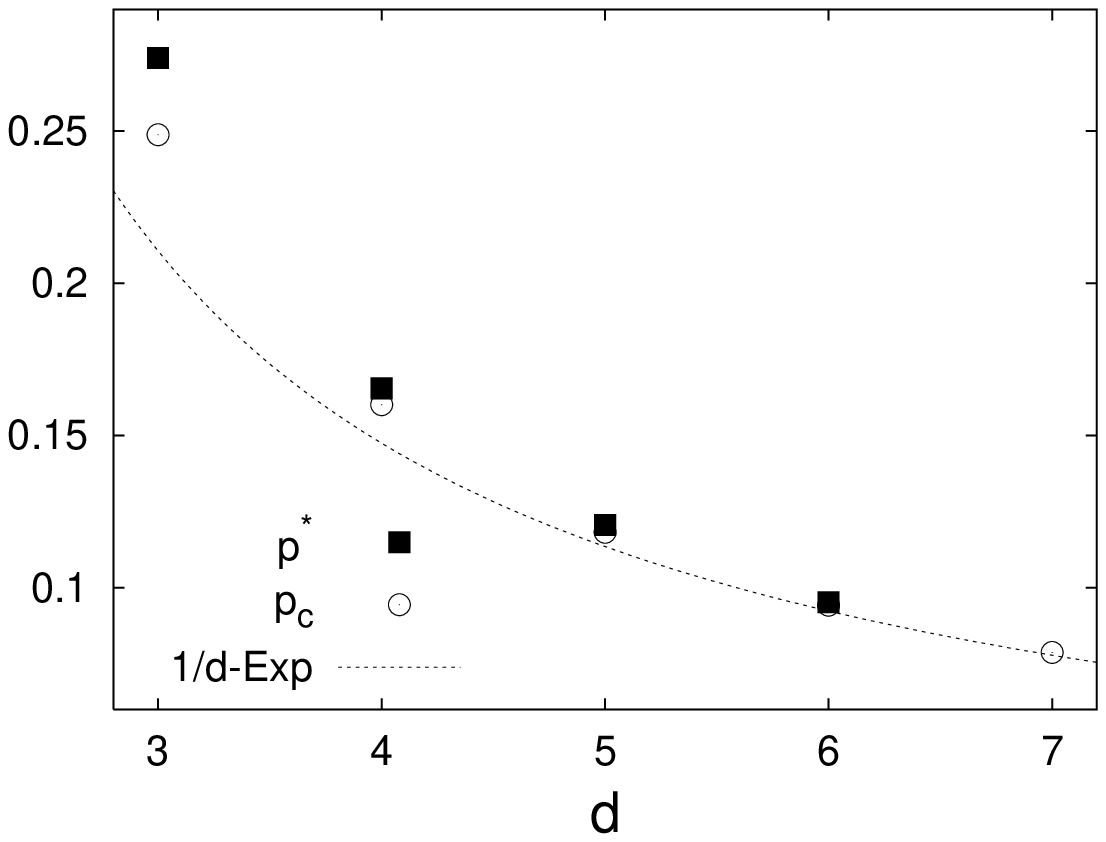}
\includegraphics{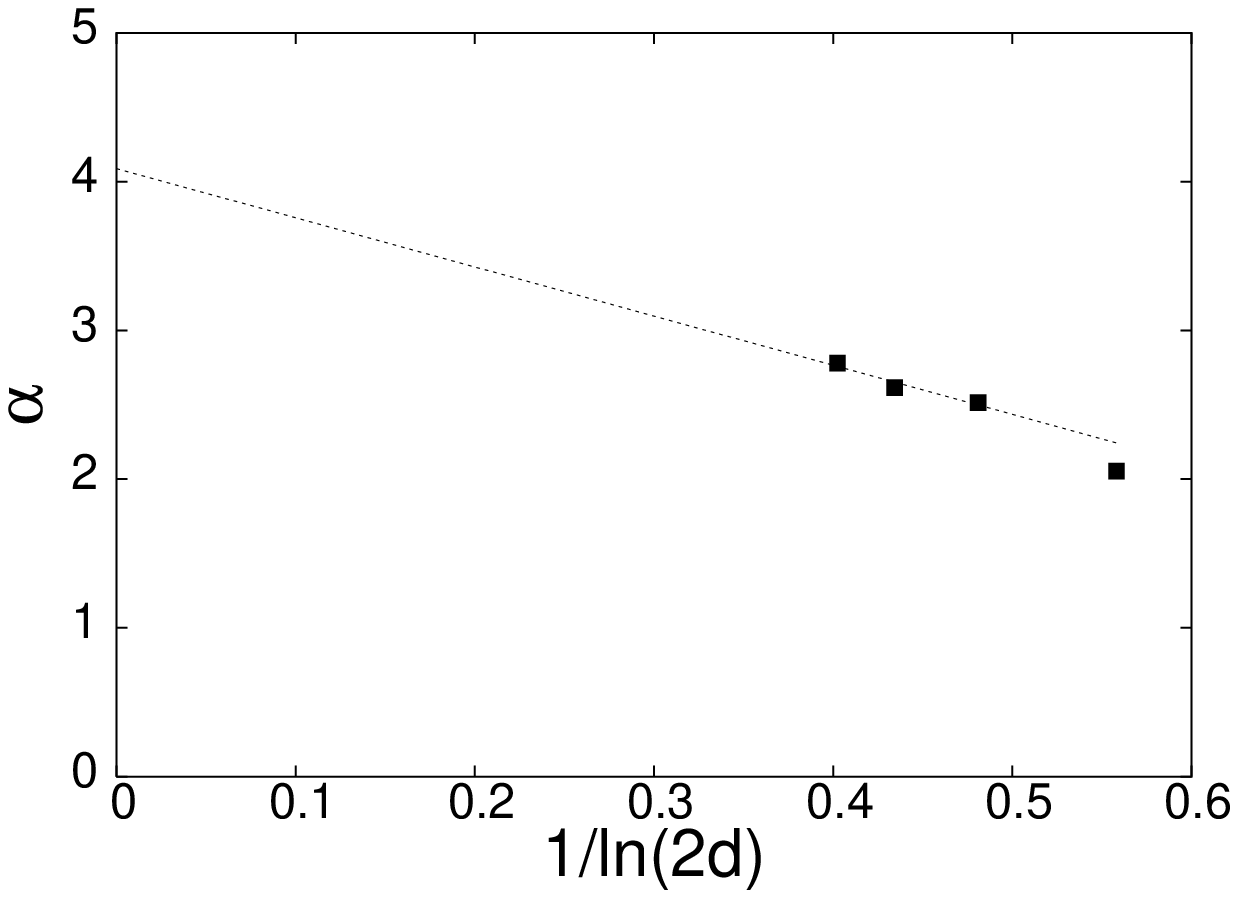}
\caption{Plot of the known bond-percolation thresholds $p_c$, the
$T=0$ glass transition thresholds $p^*$ determined in
Fig.~\protect\ref{pstarfig}, and the $1/d$-expansion of $p_c$ in
Eq.~(\protect\ref{largedeq}) as a function of $d$. The insert shows
$\ln(p^*-p_c)/\ln(2d)$ vs. $1/\ln(2d)$, extrapolating seemingly
toward $\alpha\geq4$ as in Eq.~(\ref{pscaleq}).}
\label{threshdfig}
\end{figure}

It is interesting to compare the values of $p^*$ to those of $p_c$ for
increasing dimension $d$. In Fig.~\ref{threshdfig} we plot both, $p_c$
and $p^*$, as a function of $d$, together with the prediction of the
three-term $1/d$ expansion ~\cite{Hara}
\begin{eqnarray}
p_c\sim
\frac{1}{2d}+\left(\frac{1}{2d}\right)^2+\frac{1}{2}\left(\frac{1}{2d}\right)^3
\label{largedeq}
\end{eqnarray}
The difference $p^*-p_c$ clearly decreases for $d\to\infty$. Assuming
\begin{eqnarray}
p^*-p_c \sim (2d)^{-\alpha}\quad (d\to\infty),
\label{pscaleq}
\end{eqnarray}
we plotted $\ln(p^*-p_c)/\ln(2d)$ vs. $1/\ln(2d)$ in the insert of
Fig.~\ref{threshdfig} to extrapolate for $\alpha$. This crude
extrapolation suggests $\alpha\geq4$, so that $p^*$ may share the
$1/d$-expansion of $p_c$ in Eq.~(\ref{largedeq}), at least up to the
given order. In any case, a bond-diluted lattice system with discrete $\pm J$
bonds enters its spin glass phase at an average connectivity
$2dp^*\approx1$, and the reduction methods outlined in Sec.~\ref{algo}
should be very effective in {\it any} sufficiently large dimension for
$p\gtrsim p^*$.

The value of $p^*$ is distribution-dependent~\cite{BF}, and the values
determined here and listed in Tab.~\ref{pstartable} result from
discrete $\pm J$-bonds. It is expected that $p^*=p_c$ for any
continuous distribution. The precise values for $p^*$, while
interesting in their own right, are not important for the following
discussion of the defect energy scaling. We merely need to ensure a
selection of bond densities sufficiently {\it above} $p^*$, where we
would expect Eq.~(\ref{yeq}) to hold, and sufficiently close to $p^*$
for an effective application of the reduction rules in
Sec.~\ref{algo}.

\subsection{Determination of Defect Energy Scaling}
\label{defect} 
We have conducted extensive numerical experiments to extract the
asymptotic scaling of $\sigma(\Delta E)$ for a many conveniently
chosen bond densities $p$, especially in $d=3$, but also in higher
dimensions, up to the upper critical dimension $d=6$~\cite{F+H}.
As mentioned in Sec.~\ref{stiff}, an appropriate choice of $p$ is
crucial to ensure a good compromise between maximal algorithmic
performance (for smaller $p>p^*$) and minimal scaling corrections (for
larger $p$) that maximizes the actual scaling window. While we can
estimate the effect of $p$ on the performance of our algorithm, we
have a-priori no information about scaling corrections. We will see
that scaling corrections are indeed large for $p\to p^*$. Yet, as luck
will have it, they diminish rapidly for intermediate values of $p$ and
again {\it increase} for $p\to1$ (at least in lower dimensions, where
this limit was considered).

\begin{figure}
\vskip 8in 
\includegraphics{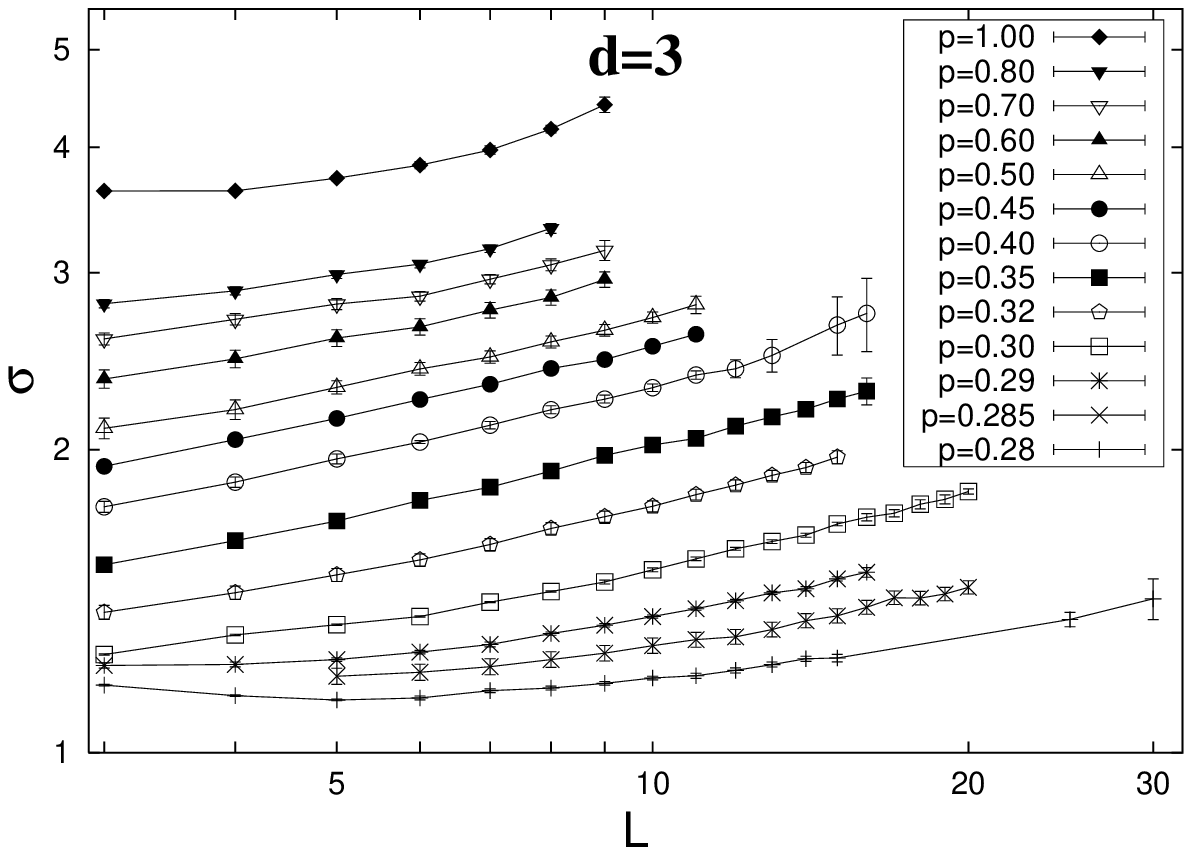}
\includegraphics{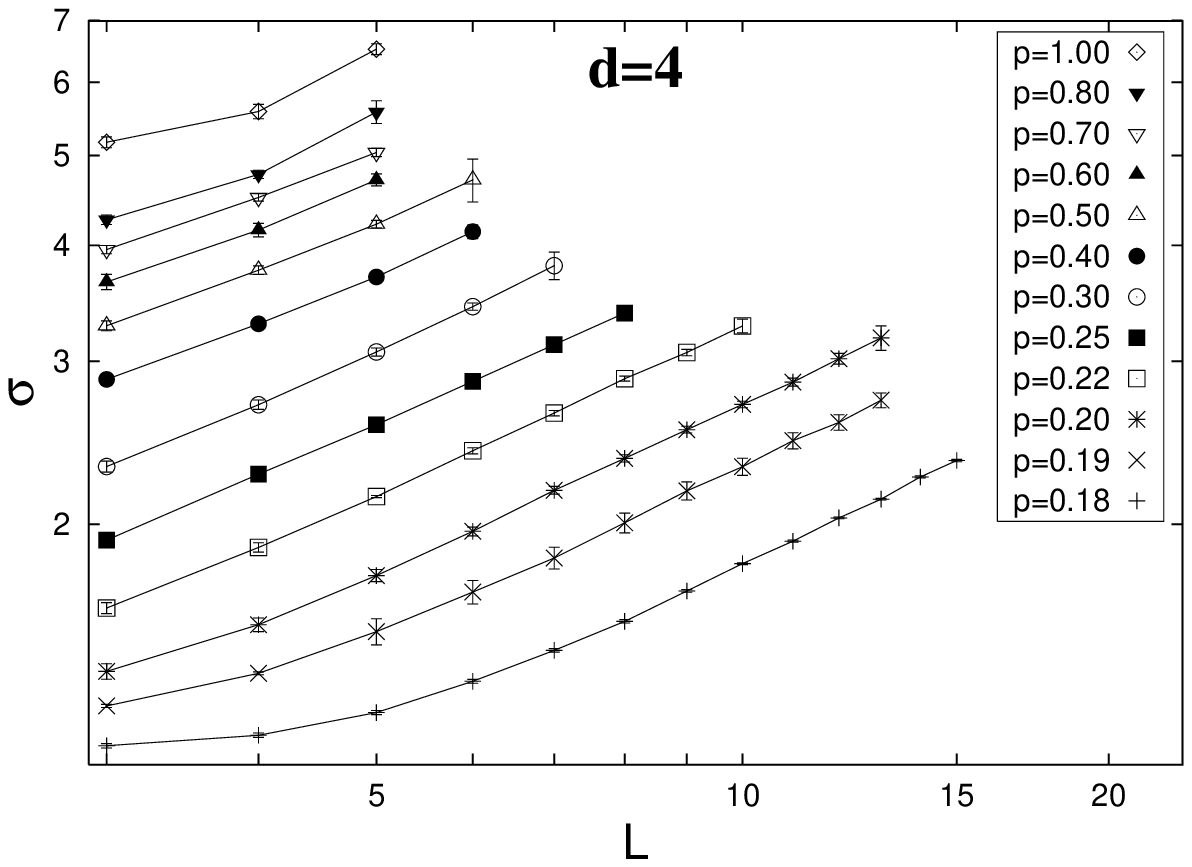}
\includegraphics{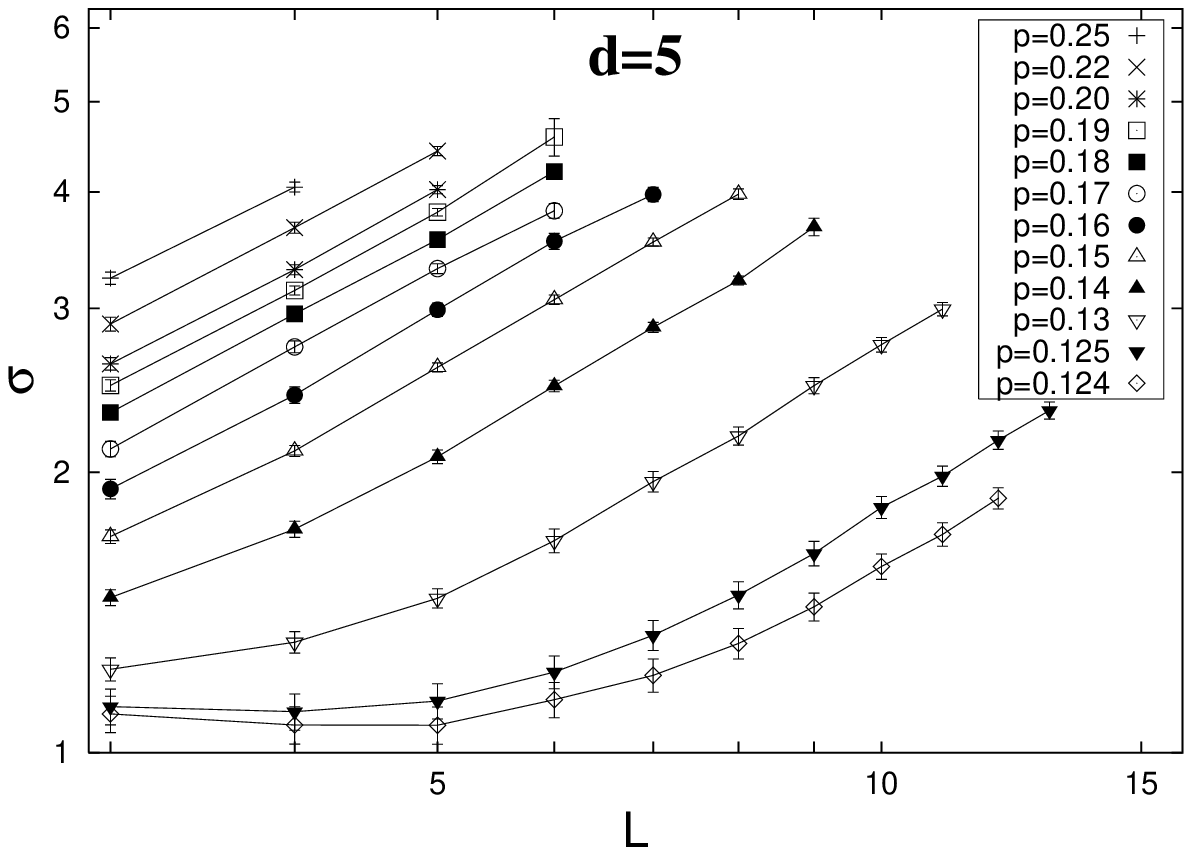}
\includegraphics{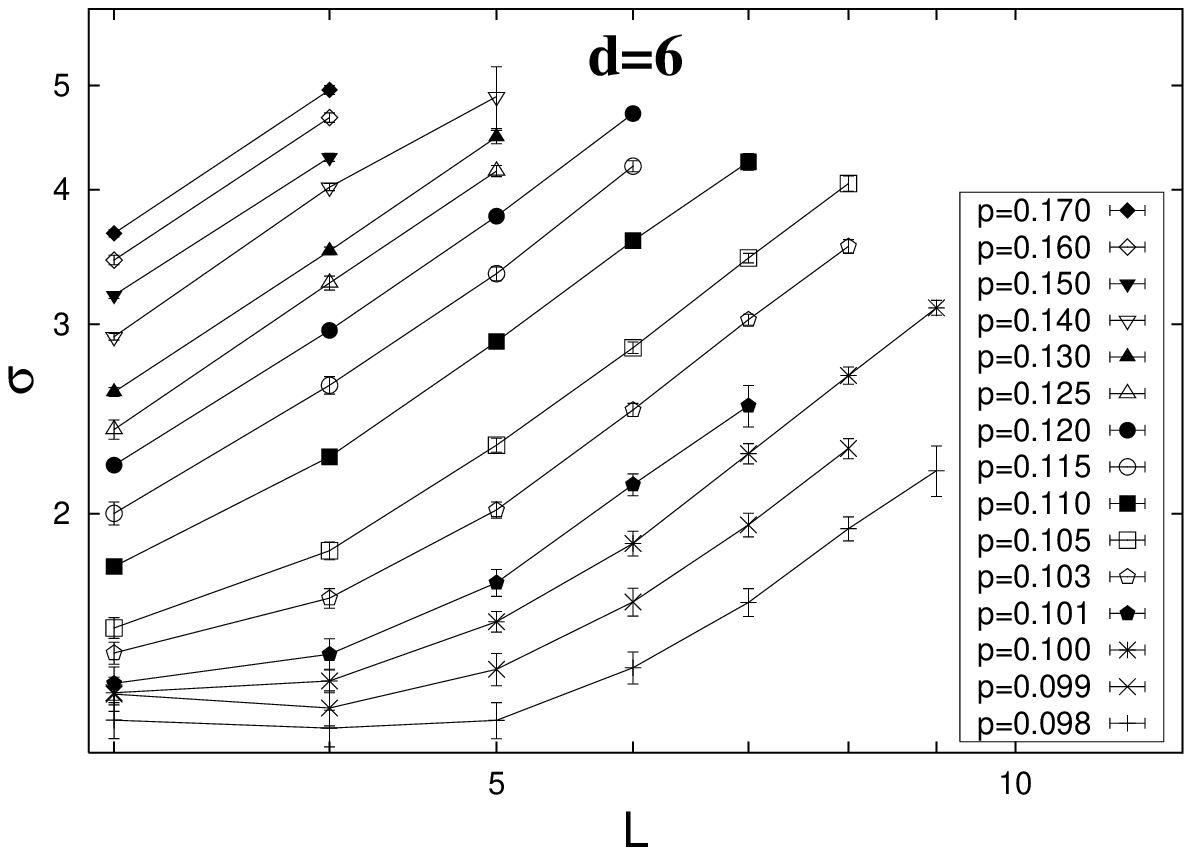}
\caption{Plot on a logarithmic scale of the width $\sigma$ of the
defect energy distribution as a
function of system size $L$.  From top to bottom, the data for
dimensions $d=3$, $d=4$, $d=5$, and $d=6$ is shown. The data is
grouped into sets (connected by lines) parameterized by the bond
density $p$. Most sets show a distinct scaling regime as indicated by
Eq.~(\protect\ref{yeq}) for a rang of $L$ above finite scaling
corrections but below failing accuracy in the optimization heuristic.
}
\label{rawdefectplot}
\end{figure}

For the study of $p^*$,  in principle very large system
sizes can be reached due to the complete reduction of very sparse
graphs. Since optimizing the spin glass on the remainder graph is an NP-hard problem, we have obtained more limited maximal system sizes above
$p^*$, dependent of the bond density $p$. We obtained sizes ranging up
to $L=30$ at $p=0.28$ to $L=9$ at $p=1$ in $d=3$, $L=15$ at $p=0.18$
to $L=5$ at $p=1$ for $d=4$, $L=13$ at $p=0.125$ to $L=5$ at $p=0.22$
in $d=5$, and $L=9$ at $p=0.1$ to $L=4$ at $p=0.17$ in $d=6$.  For
each choice of $L$ and $p$, we have sampled the defect energy
distribution with at least $N\geq10^5$ instances, then determined its
variance $\sigma(\Delta E)$. For each data point for $\sigma(\Delta
E)$ we estimated its error bar as $7/\sqrt{N}$. In
Figs.~\ref{rawdefectplot}, we plot all the data for each dimension
simply according to Eq.~(\ref{yeq}), on a logarithmic scale. For most
sets of graphs, a scaling regime (linear on this scale) is
visible. Yet, various deviations from scaling can be
observed. Clearly, each sequence of points should exhibit some form of
finite size corrections to scaling for smaller $L$. For large $L$, the
inability to determine defect energies correctly (according to the
discussion in Sec.~\ref{stiff}), will inevitably lead to a systematic
bias in $\sigma$. Some data sets did not exhibit any discernible
scaling regime whatsoever, most notably our data set for the undiluted
lattice in $d=3$.

\begin{figure}
\vskip 7.85in 
\includegraphics{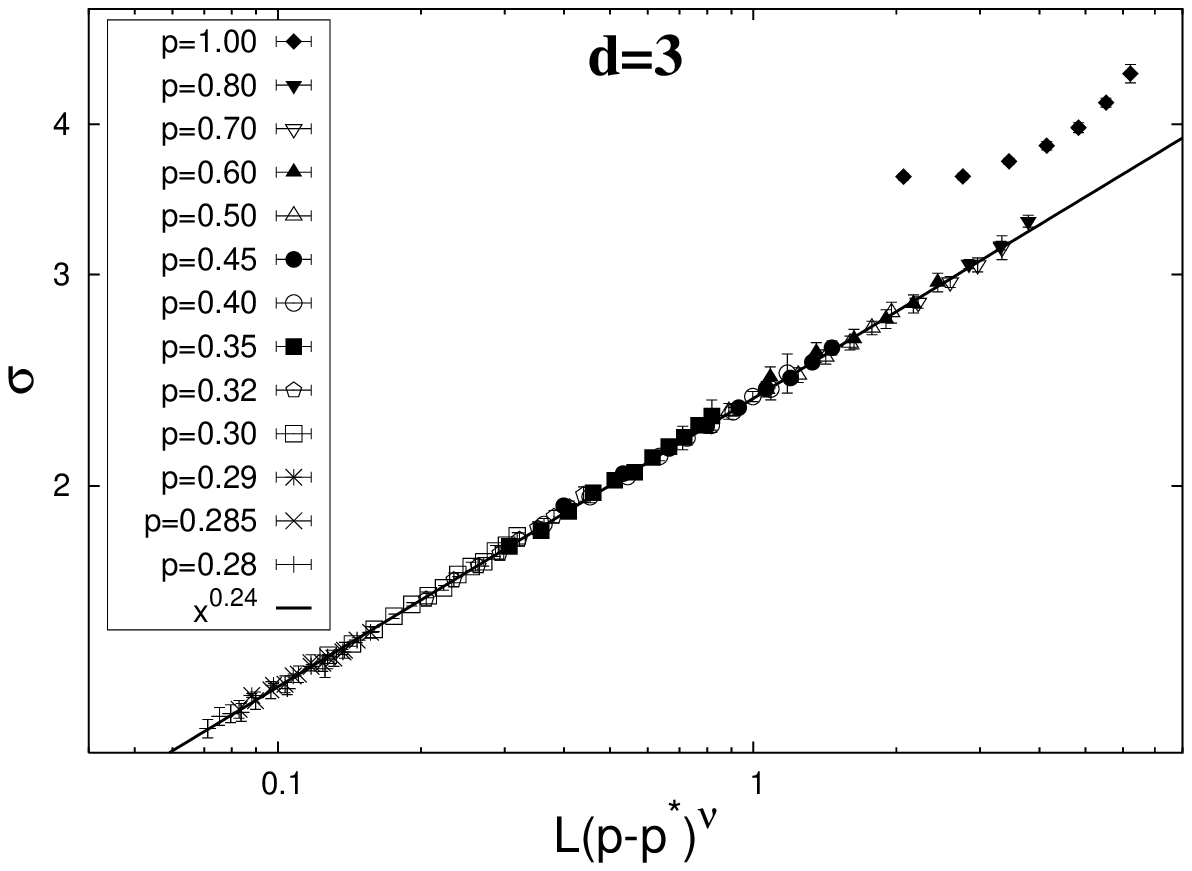}
\includegraphics{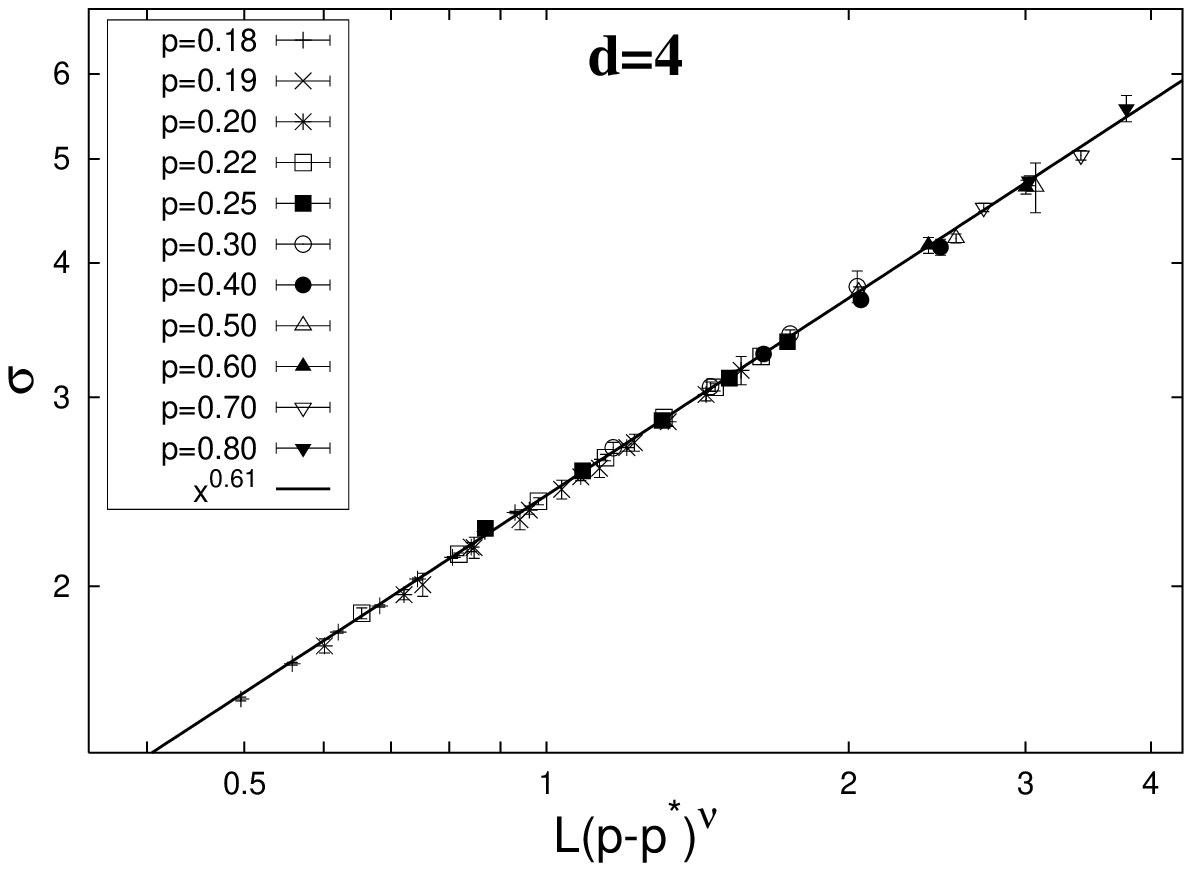}
\includegraphics{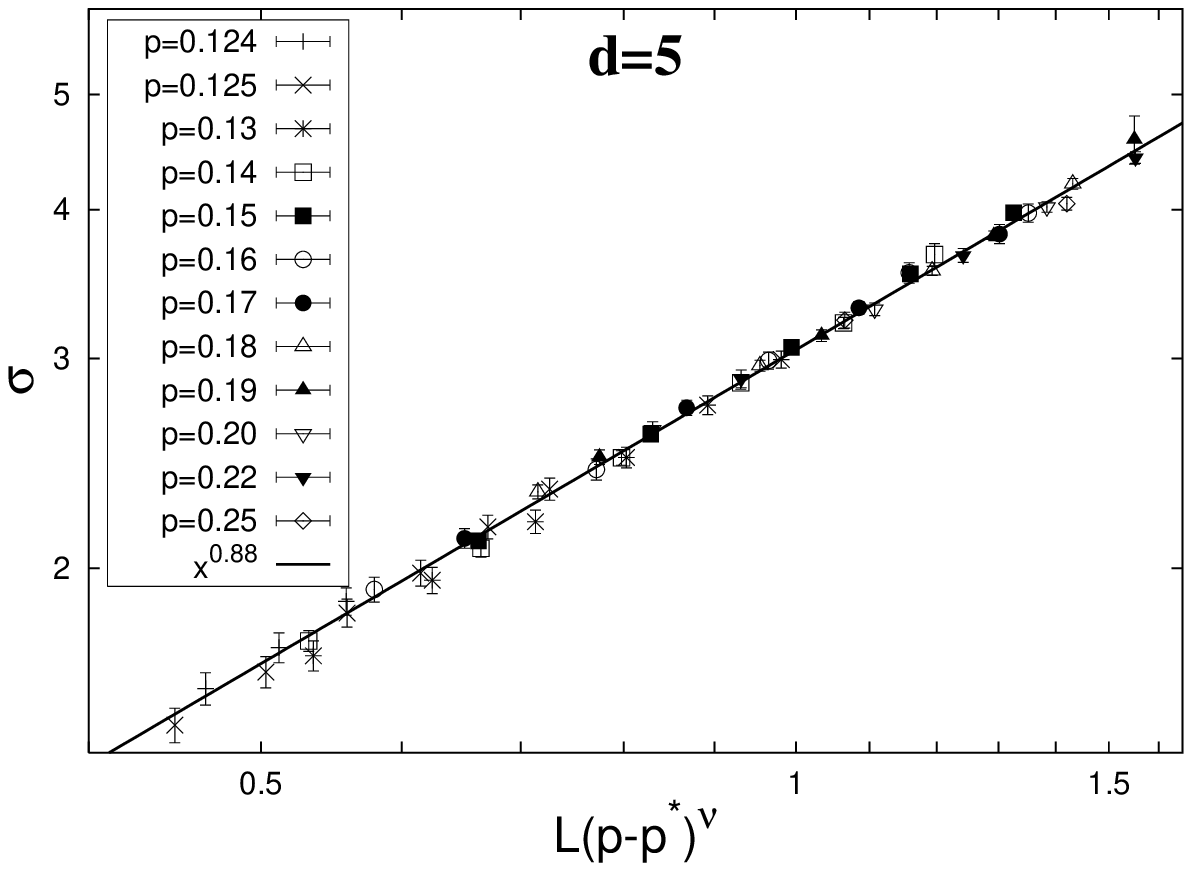}
\includegraphics{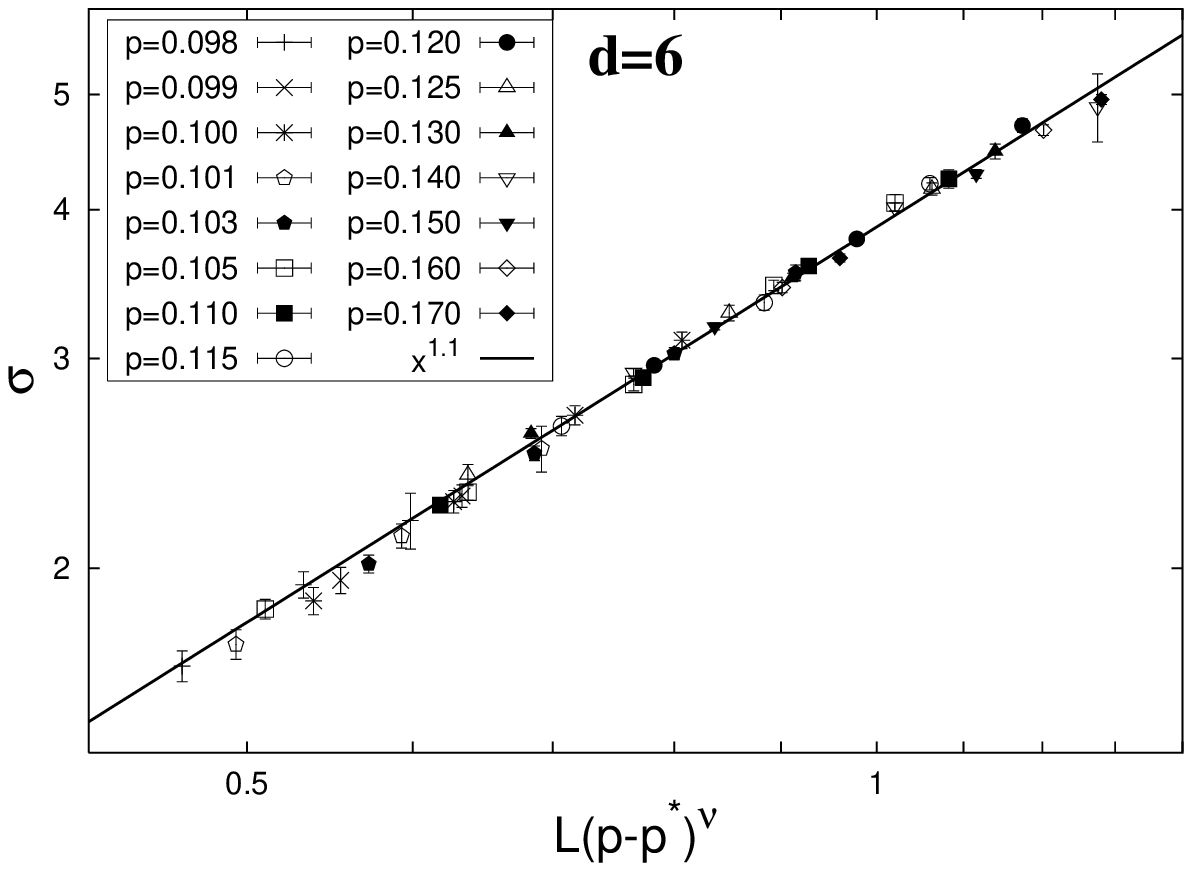}
\caption{Scaling plot of the data from
Figs.~\protect\ref{rawdefectplot} for $\sigma$, fitted to Eq.~(\protect\ref{fiteq}) as a function of the
scaling variable $x=L(p-p^*)^{\nu^*}$. Data above or below the
scaling regime in each set from Figs.~\protect\ref{rawdefectplot} was
cut. From top to bottom, the scaling collapse of the data for
dimensions $d=3$ to 6 is shown. The  lines represent a power-law fit of the collapsed data which
provides an accurate determination of the stiffness exponent $y$ in
each dimension. For $d=3$ (top), we have also included the data for 
$p=1$, which does not appear to connect to the scaling regime. }
\label{scaldefectplot}
\end{figure}

To obtain an optimal scaling collapse of the data, we focus on the
data inside the scaling regime for each set. To this end, we chose for
each data set a lower cut in $L$ by inspection. An appropriate
high-end cut is introduced by eliminating all data points for which
the remainder graph had a typical size of $>700$ spins; at that point
the EO heuristic (within the supplied runtime) seems to fail in
determining defect energies with sufficient accuracy. All the
remaining data points for $L$ and $p$ are fitted to a four-parameter
scaling form,
\begin{eqnarray}
\sigma(\Delta E)\sim {\cal Y}_0\,
\left[L\left(p-p^*\right)^{\nu^*}\right]^y,
\label{fiteq}
\end{eqnarray}
i.~e. approximating the scaling function $g(x)$ from
Eq.~(\ref{newyeq}) merely by unity, its leading behavior for large argument. Unfortunately, we have no knowledge of the functional form for
finite-size corrections, making the low-$L$ cut on the data a
necessity. The fitted values for this and the other fitting constants
($p^*$, $\nu^*$, and $y$) are listed in
Tab.~\ref{fitdatatable}. Holding $p^*$ fixed at the independently
determined values from Tab.~\ref{pstartable} reduces the variance in
the remaining fitting parameters without changing much in their quoted
values. Using the parameters of this fit, we re-plot only the data from
the scaling regime in each dimension in Figs.~\ref{scaldefectplot}.

In each case, a convincing scaling collapse is obtained. Clearly, our
data for $d=3$ is not only the most extensive, but also happens to
scale over nearly two decades without any discernible deviation or
trend away from pure power-law scaling that would betray any
systematic bias or lack of asymptotic behavior. This justifies a
certain degree of confidence to project $y_3=0.24(1)$ for the scaling
exponent where the quoted error is based on the uncertainty in the
fit. Troublesome is the observation that the data for the undiluted
lattice ($p=1$) never reaches the scaling regime (see
Figs.~\ref{scaldefectplot}, top). This may be in accordance with the
observation of Ref.~\cite{DM}, which found very long transients in a
similar study on undiluted Migdal-Kadanoff lattices (see also
Ref.~\cite{MKpaper}), or similar findings for undiluted lattices~\cite{Middle}. In our data, systematic errors in sampling ground states seem to set in for large $L$, before any scaling regime is reached at all.

For increasing dimension $d$, accessible scaling regimes become
shorter, leading to more difficulty in determining an accurate fit of
the power law. In $d=4$ we can still claim scaling for about a decade
in the scaling variable, justifying a prediction of $y_4=0.61(2)$. In
$d=5$ and $d=6$, we only reach scaling windows significantly
shorter than a decade. Luckily, $y_d$ increases with increasing $d$, thus
larger absolute errors still result in acceptable relative errors, and
we predict from the fits in Figs.~\ref{scaldefectplot} that
$y_5=0.88(5)$ and $y_6=1.1(1)$.

\begin{table}
\caption{List of the fitted values for the critical bond-density
$p^*$,the correlation-length exponent $\nu^*$, the surface tension
${\cal Y}_0$, and the stiffness exponent $y$ in each dimension $3\leq
d\leq6$. Included is also the $Q$-value for each fit. The values for
$p^*$ here are bound to be less accurate than those directly
determined in Sec.~\protect\ref{pstar}, but are consistent. In
contrast, the values for $y$ are quite stable.}
\begin{tabular}{r|lllll}
\hline\hline 
$d$ &  $p^*$  & $\nu^*$ & ${\cal Y}_0$ &  $y$  & $Q$~(DoF)\\ 
\hline 
3   & 0.2706  & 1.17    & 2.37         & 0.239 & 1.00~(92) \\ 
4   & 0.1699  & 0.60    & 2.43         & 0.610 & 0.00~(47) \\ 
5   & 0.1217  & 0.50    & 3.05         & 0.876 & 0.86~(48) \\ 
6   & 0.0959  & 0.44    & 3.87         & 1.103 & 0.02~(46)  \\ 
\hline\hline
\end{tabular}
\begin{tabular}{r|lllll}
\hline\hline 
\end{tabular}
\label{fitdatatable}
\end{table}

\section{Conclusions}
\label{conclusion}
We have used the combined effort of an exact reduction method and an
efficient heuristic to determine the defect energy distribution for
$\pm J$-spin glasses on bond-diluted lattices in low dimensions. A
subsequent finite size scaling fit of the data allowed us to extract
the stiffness exponents in these dimensions to within 4\% to 10\%
accuracy. Our approach also allowed the determination of a variety of
other observables associated with the $T=0$ transition into a glassy
state at a bond-density $p^*$ for $d\geq3$. We hope that the methods
introduced here may be applicable as well to the treatment of other
open questions regarding the low-temperature state of spin glass
systems~\cite{KM,PY2}.

Our value of $y_3=0.24(1)$ in $d=3$ is near the higher end of previous
estimates varying between $0.19$ and $0.27$, while $y_4=0.61(2)$ in
$d=4$ overlaps with a previous result of $0.64(5)$ from
Ref.~\cite{Hd4}. There has been no previous value for $d=5$, but for
$d=6$, the upper critical dimension, our value of $y_6=1.1(1)$ suggest
a mean field result of $y_{\infty}=1$. There has been no previous
determination of the exponents $\nu^*$, except that it is bound to
exceed the value of $\nu$ for percolation \cite{BF}, and that its mean
field value for $d\geq6$ should be $\nu^*_{\infty}=1/2$. In light of
the fact that $\nu=0.875,$ 0.68, 0.57, and 0.5 for bond percolation in
$d=3$ to 6~\cite{AMAH}, most of our fitted values for $\nu^*$ do not seem to
satisfy these expectations, which is easily explained with
their poor accuracy. For instance, $\nu^*\geq1/2$ should hold, so it
appears that the fitted values of $\nu^*$ are generally too low.

I would like to thank A. Percus, R. Palmer, M. Palassini, and
A. Hartmann for helpful discussions, and our IT staff for providing
access to our student computing lab. This project is supported by
grant 0312510 from the Division of Materials Research at the National
Science Foundation.

\end{document}